\documentclass[11pt,preprint,graphicx]{aastex}
\usepackage{graphicx}

\def\etal{\it et al. \rm }

\begin{document} 

\title{The Structure of Galaxies: II. Fitting Functions and
Scaling Relations for Ellipticals}

\author{J. M. Schombert$^{A,B}$}
\affil{$^A$Department of Physics, University of Oregon, Eugene, OR USA 97403}
\affil{$^B$jschombe@uoregon.edu}

\begin{abstract}

\noindent Surface photometry of 311 ellipticals from the 2MASS imaging
database is analyzed with respect to the two most common fitting
functions; the $r^{1/4}$ law and the S\'{e}rsic $r^{1/n}$ model.  The
advantages and disadvantages of each fitting function are examined.  In
particular, the $r^{1/4}$ law performs well in the middle regions, but is
inadequate for the core (inner 5 kpcs) and the outer regions (beyond the
half-light radius) which do not have $r^{1/4}$ shapes.  It is found that
the S\'{e}rsic $r^{1/n}$ model produce good fits to the core regions of
ellipticals ($r < r_{half}$), but is an inadequate function for the entire
profile of an elliptical from core to halo due to competing effects on the
S\'{e}rsic $n$ index and the fact that the interior shape of an elliptical
is only weakly correlated with its halo shape.  In addition, there are a
wide range of S\'{e}rsic parameters that will equally describe the shape of
the outer profile, degrading the S\'{e}rsic models usefulness as a
describer of the entire profile.  Empirically determined parameters, such
as half-light radius and total luminosity, have less scatter than fitting
function variables.  The scaling relations for ellipticals are often
non-linear, but for ellipticals brighter than $M_J < -23$ the following
structural relations are found: $L \propto r^{0.8 \pm 0.1}$, $L \propto
\Sigma^{-0.5 \pm 0.1}$ and $\Sigma \propto r^{-1.5 \pm 0.1}$.

\end{abstract}

\section{Introduction}

The structure of elliptical galaxies, as inferred from surface brightness
profiles, is the most direct method of deriving the size, luminosity and
density scale parameters that are key to understanding the details of
galaxy formation.  This type of information has become increasing important
as our successful $\Lambda$CDM cosmological simulations begin to focus on
smaller scale cluster and galaxy sized predictions (Tonini \etal 2010,
Trujillo-Gomez \etal 2011).  Current formation scenarios range from
gravitational collapse on short timescales to extended structure evolution
by mergers of gas-rich (wet) and gas-poor (dry) companions in a
hierarchical fashion (Steinmetz \& Navarro 2002).  Determining the
characteristics of structure in present-day galaxies is also a critical
step to understanding structural changes at high redshift (Chevance \etal
2012).

Interpretation of surface photometry commonly uses fitting functions, which
were introduced to surface brightness profiles to provide parametrization
after it was discovered that ellipticals varied in structure in a uniform
fashion with size or luminosity.  A simple set of parameters would allow
for quantitative classification of ellipticals and the identification of
structure components that might be related to kinematic properties.  In
addition, describing structure with fitting functions provides an avenue to
locate evolutionary signatures (such as mergers, dust lanes or tidal
interactions) and allows for comparison to theoretical predictions of
galaxy structure (Mosleh \etal 2013).  Ultimately, uniform structure
described by a simple function implies homology for galaxy formation
(Bertin \etal 2002) with the hope of revealing a universal profile shape
that reflects the underlying baryonic and dark matter distributions
(Navarro \etal 1997, Merritt \etal 2006), although similarity may be a
function of both structure and kinematics (Navarro \etal 2010).

The mechanical goal of fitting functions is to reduce the 2D shape of the
surface brightness profile to a set of simple parameters that are
mathematically related.  This would, in effect, allow for the complete
reconstruction of the luminosity density of a galaxy from a small set of
values.  However, simply finding a spline-like function that matches all
the data points is inadequate for a description of a profile as it would
have too many variables and does not allow meaningful comparison of those
values with other photometric or kinematic properties of galaxies.  The
mathematically simplest formula is expected to be the one that provides the
greatest correlation between structural and photometric characteristics
and, therefore, revealing more of the underlying physics.

The history of fitting functions is tied to the technological progress of
galaxy photometry from the early days of photographic plates to the advent
of electronic detectors (e.g., CCD's).  Through the infancy of galaxy
photometry, the fitting functions for ellipticals progressed from the
Reynold's (1913) model, to Hubble's (1930), a modified Hubble (Rood \etal
1972) and lastly a truncated Hubble model (Oemler 1976) (see Graham 2011
for a complete review).  Parallel to these efforts, which focused on the
halo fits (the region beyond the half-light radius) in order to reveal mass
density, was the $r^{1/4}$ law developed by de Vaucouleurs (1953) primarily
to confine the curves of growth for aperture photometry.

The $r^{1/4}$ surface brightness law, as first outlined by de Vaucouleurs
(1948), was first reinforced as the fitting function of choice by its
excellent representation of the deep surface photometry of NGC 3379 (de
Vaucouleurs \& Capaccioli 1979).  While shown to be inadequate for dwarf
ellipticals, the popularity of the $r^{1/4}$ law continued into the 1980's
to the point where it was considered a universal fit to all ellipticals,
and deviations from a $r^{1/4}$ fit were interpreted as the result of tidal
interactions (Kormendy 1977).

The universality of the $r^{1/4}$ law was questioned with the discovery
that the its two variables, effective radius ($r_e$) and surface brightness
($\mu_e$), were coupled and decreased the meaning of their correlations
(Kormendy 1980, Schombert 1986).  In addition, it was shown in Schombert
(1987), that ellipticals deviated from the $r^{1/4}$ law in a systematic
fashion with luminosity.  Clearly, two parameters were insufficient to
adequately describe the structure of ellipticals over a full range of
luminosities, even excluding dwarfs and giant cD galaxies (Schombert 1987).

The need for addition parameters to capture additional shape beyond the
$r^{1/4}$ law resulting in the adoption of the S\'{e}rsic (1963)
generalization, a $r^{1/n}$ model, where effective radius and surface
brightness are joined by a concentration variable, $n$.  This fitting
function has the immediate advantage in that the S\'{e}rsic $r^{1/n}$ model
runs from exponential (i.e., $n = 1$, well suited for disk galaxies and
dwarf ellipticals) to $r^{1/4}$ (i.e. $n = 4$) and higher values of $n$ for
brighter luminosity ellipticals.  Another benefit of the S\'{e}rsic
$r^{1/n}$ model was its application as a photometric plane for ellipticals
(Graham 2002), an analogous relation to the Fundamental Plane.  Extensions
of the S\'{e}rsic $r^{1/n}$ model are used to interpret high resolution
space imaging (Graham 2005), but our study focuses only on the outer
regions of ellipticals.

The goal of this paper, the second in our series on the structure of
galaxies, is examine the usefulness of fitting functions in describing the
outer isophotes of ellipticals.  The success in the S\'{e}rsic $r^{1/n}$
model for parameterizing the core regions (those regions inside the
half-light radius, typically between 4 and 6 kpcs) of ellipticals is well
established (Graham \& Guzman 2003).  However, a majority of those studies
focus on the inner isophotes, at the sacrifice of information from the
halo.  In this paper, usefulness of both the $r^{1/4}$ law and the
S\'{e}rsic $r^{1/n}$ model to the halos of ellipticals will be examined,
and what scaling relations can be extracted for this most common type of
galaxy in rich, dense environments.

\section{Data}

As described in Paper I (Schombert \& Smith 2012), the images for this
study are taken from the 2MASS Image archive.  The 2MASS project was a NASA
ground-based, all-sky, near-IR sky survey (Skrutskie \etal 2006).  2MASS
uniformly scanned the entire sky using two 1.3-m telescopes (north KPNO and
south CTIO).  Each telescope was equipped with a three-channel camera,
where each channel consisting of a 256x256 HgCdTe detector.  Each camera
was capable of observing the sky simultaneously at $J$ (1.25 microns), $H$
(1.65 microns), and $K$ (2.17 microns).  The 2MASS arrays imaged the sky in
a drift-scan mode.  Each final pixel consisted of six pointings on the sky
for a total integration time of 7.8 sec per pixel.  The final image frames
have a plate scale of one arcsec per pixel with typical depth of 24 $J$ mag
arcsecs$^{-2}$ (errors at 0.5 mags).  

The sample was selected by morphological criteria from the Revised
Shapley-Ames (RSA) and Uppsala Galaxy Catalogs (UGC).  All the galaxies
must be pure 'E' classification in both catalogs.  In addition, the
selected galaxies had to be free of nearby companions or bright stars which
might disturb the analysis of the isophotes to faint luminosity levels and
sufficiently small in angular size to cover only two 2MASS strips.  The
final sample contained 428 galaxies and covers apparent $J$ magnitudes from
7 to 11.5 and absolute $J$ magnitudes from $-$21 to $-$26.  In the process
of reducing the surface brightness profiles, it was found that the galaxies
divided into two subsamples that will be discussed in Paper III.  For this
study, 311 clean ellipticals with clear single component profiles were
isolated.

Images from 2MASS for regions around all the galaxies in the sample were
downloaded from 2MASS's Interactive Image Service.  These sky images were
flattened and cleaned by the 2MASS project and contained all the
information needed to produce calibrated photometry.  The images were
analyzed as described in \S 3 of Paper I.  All the reduced photometry can be
found at our data website (http://abyss.uoregon.edu/ $\sim$js/sfb).

\section{$r^{1/4}$ fits}

Since the $r^{1/4}$ law was the fitting function of choice for many
decades, this function was fit to all the galaxies in our sample.  The
shortcomings to the $r^{1/4}$ law is well documented in Graham (2011) and,
in particular, it was shown by Schombert (1986) that ellipticals are only
$r^{1/4}$ in shape for a very limited range of surface brightness
(typically between 21 and 24 $V$ mag arcsecs$^{-2}$) and for a limited
range in total luminosity (i.e., galaxies less than $M_V = -20.5$ have no
portion of their surface brightness profiles which are $r^{1/4}$ in shape).

Following the prescription of Schombert (1986), only that portion of the
surface brightness profile which is linear when plotted in $r^{1/4}$ space
is fit.  This can be done in a subjective manner by visually selecting the
inner and outer radii in a plot of $\mu$ versus $r^{1/4}$, or can be
automated by restricting the fit to between 19 and 22 $J$ mag
arcsecs$^{-2}$ and searching for the best linear region.  Either method
produces identical results in terms of similar structural correlations.

\begin{figure}[!ht]
\centering
\includegraphics[scale=0.75,angle=0]{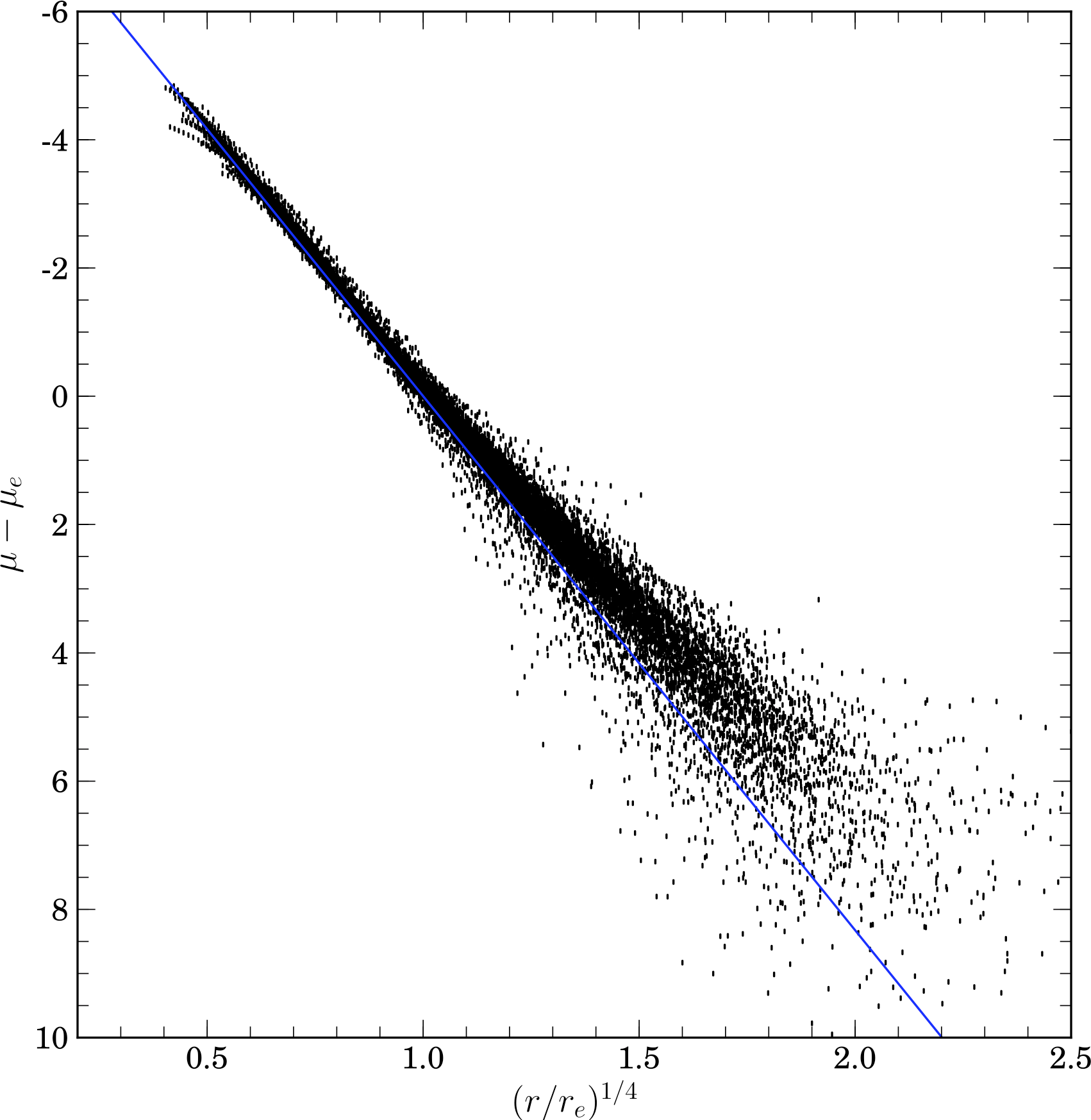}
\caption{\small The surface brightness profiles of all 311 ellipticals in
our sample normalized to their best $r^{1/4}$ fit.  The blue line indicates
the exact $r^{1/4}$ shape, and it is clear that most ellipticals deviate
above the $r^{1/4}$ law at large radii and that the $r^{1/4}$ shape fails
for the inner regions ($r < 2$ kpcs).  However, despite it's limitations
for outer isophotes, the $r^{1/4}$ shape is so consistent for the middle
regions that this fact must be address by any structural model.
}
\label{normal_re}
\end{figure}

Figure \ref{normal_re} displays all the galaxies in our sample, normalized
for their best $r^{1/4}$ fit.  Only data at radii greater than 2 arcsecs
are displayed to avoid seeing effects.  The deviations from the $r^{1/4}$
are clear to see in this figure, being typically higher in surface
brightness at large radii than the $r^{1/4}$ law for bright galaxies, less
than the $r^{/4}$ law for faint galaxies.  However, for the restricted
range of surface brightness, the $r^{1/4}$ law is a good description of the
interior structure of ellipticals.

It is surprising that the arbitrary nature of the fitting process results
in similar structural relations (e.g. Figure \ref{re_se}).  However, this
is due to the coincidence of interior versus outer structure in ellipticals
as compared to the $r^{1/4}$ law.  Ellipticals, typically, will have some
downward turn in surface brightness in their interior regions due to having
interior structure following a S\'{e}rsic model with $n < 10$ (Graham
2011).  Likewise, there is an upward turn in surface brightness at outer
radii as can be seen in Figure \ref{normal_re}.  This will result in a
natural bias towards steeper slopes as one includes interior and exterior
data.  PSF effects can also contribute to this problem and, as shown in
Paper I, 2MASS images have measurable PSF distortions out to 4 arcsecs.
PSF errors will distribute core luminosity outward, producing a shallower
slope for inner isophotes.

\begin{figure}[!ht]
\centering
\includegraphics[scale=0.8,angle=0]{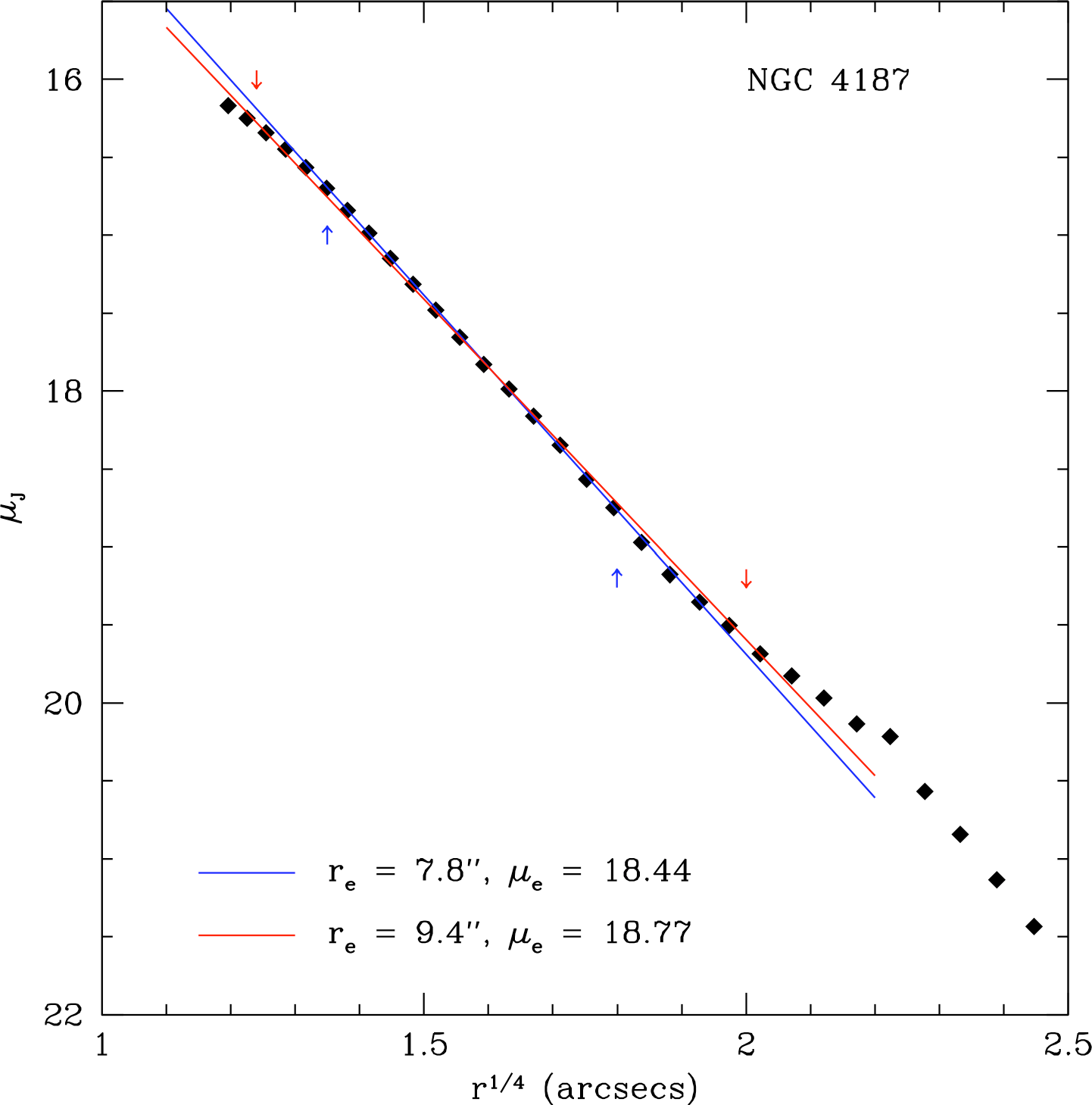}
\caption{\small An example of the difficulty in finding correct $r^{1/4}$
fits for most ellipticals.  The typical behavior for an elliptical profile
is to curve fainter towards the core and brighter in the halo.  This
results in a subjective decision on which isophotes to use for fitting.
The two ends drive $r_e$ and $\mu_e$ to larger values, although in such a
fashion as to preserve the photometric $\mu_e$-log $r_e$ relation.  The
arrows indicate the range of isophotes used for each fit.
}
\label{ngc4187}
\end{figure}

An illustration of this effect is seen in Figure \ref{ngc4187}.  Plotted is
the surface brightness profile of NGC 4187 in $r^{1/4}$ space.  A straight
line is a good match to the $r^{1/4}$ law, as is shown by the blue line
(fit range indicated by blue arrows).  However, a formal fit that includes
only a few more interior and exterior points (the red line and red arrows)
results in a fit that is 20\% larger in effective radius ($r_e$) and an
effective surface brightness ($\mu_e$) that is 35\% fainter.  When previous
studies referred to the coupling of $r^{1/4}$ parameters (Trujillo, Graham
\& Caon 2001), it is this effect that causes the coupling.  Notice that the
bias in $r_e$ and $\mu_e$ results in the change in the measured structural
parameters that is nearly parallel to the overall relationship between
$r_e$ and $\mu_e$ (the errors in the fit produce a $\Delta \Sigma \propto
\Delta r^{-2}$, where the relation in Figure \ref{re_se} is $\Sigma_e
\propto r_e^{-3}$), and is one of the main reasons the scatter is so small
over such a large range in galaxy size and luminosity.

The resulting structural scaling relation, log $r_e$ versus $\mu_e$, is
shown in Figure \ref{re_se}.  A jackknife linear fit gives $\mu_e$ = $2.99
\pm 0.04$ log $r_e$ + $16.95 \pm 0.02$.  Also shown in the Figure is the
relationship from Kormendy (1977), corrected to an $H_o = 72$ and a $B-J$
color of 3.5 ($\mu_e$ = 3.28 log $r_e$ + 16.77).  The outliers with small
$r_e$ and faint $\mu_e$ values are galaxies where, even with fitting
restrictions, are not well fit by the $r^{1/4}$ law in any region of their
surface brightness profile.  The correlation is real from the UV to the
near-IR, but the low scatter is, in some part, due to the coupling of the
fit parameters.  The structural values for a particular galaxy is much more
uncertain than indicated by the tightness of the correlation.

\begin{figure}[!ht]
\centering
\includegraphics[scale=0.8,angle=0]{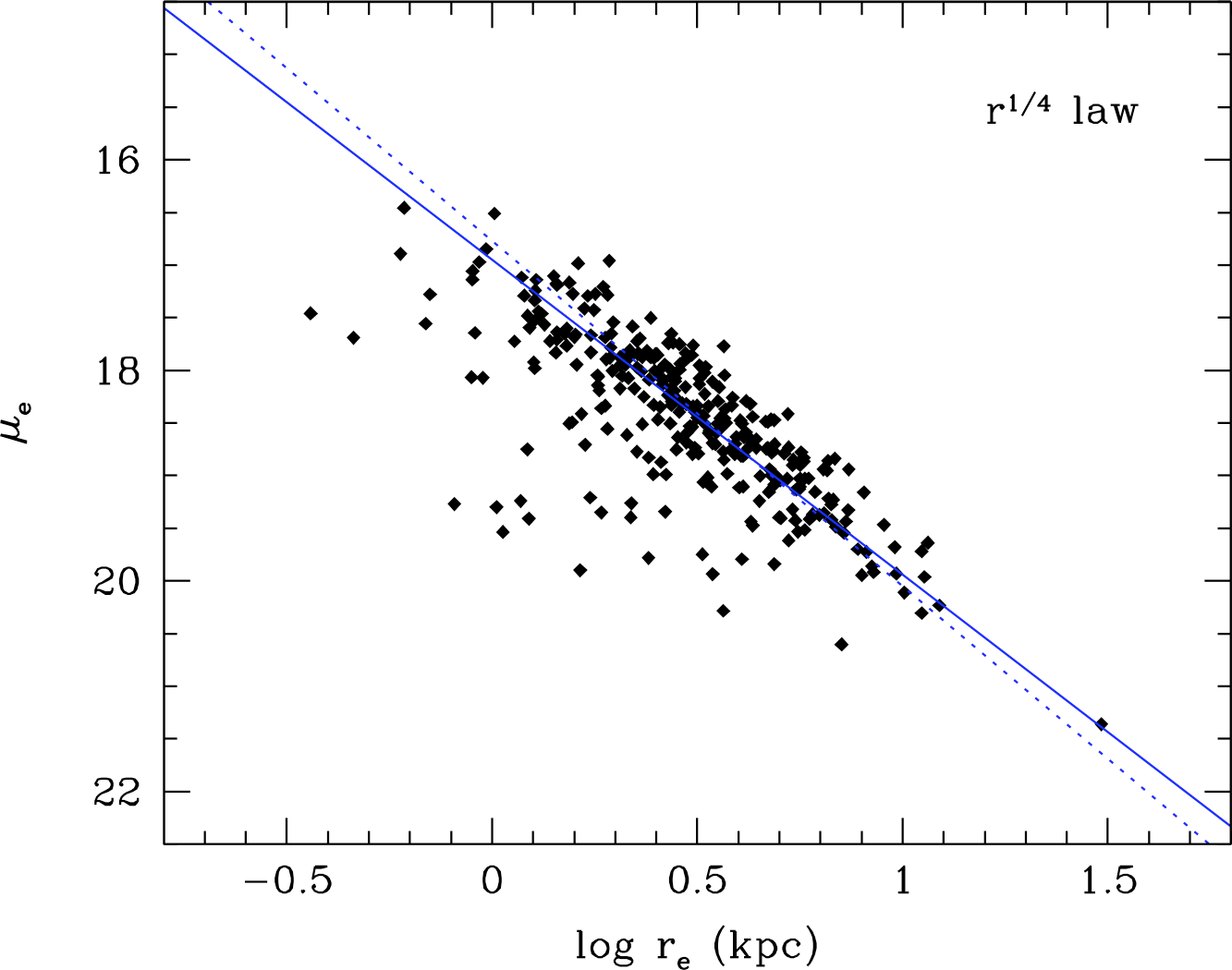}
\caption{\small The correlation between effective radius ($r_e$) and
effective surface brightness ($\mu_e$) for $r^{1/4}$ fits.  The blue line
is a jackknife linear fit, the dashed line is the relation from Kormendy
(1997) corrected for a mean $B-J$ color of 3.5.  Despite different fitting
techniques, three decades in time and 5000\AA\ in wavelength, the same
relationship is found for the 2MASS sample as the Kormendy sample.
}
\label{re_se}
\end{figure}

The quality of the correlation underlies the success of the $r^{1/4}$ law
for many years.  For, even though the $r^{1/4}$ systematically fails to fit
the outer portion of ellipticals, it does fit the middle portions where a
majority of the light is located.  The two fit variables give a crude map
of the galaxy shape and correlate with various global parameters, such as
total luminosity.  So the $r^{1/4}$ law, although it fails as a descriptor
of central concentration and halo extension, does serve as a basic
indicator of mean galaxy size and luminosity density.  But the inclusion of
low luminosity ellipticals, which have no region of their profiles which
are $r^{1/4}$ in shape, will destroy this relationship.

\section{S\'{e}rsic $r^{1/n}$ model}

\begin{figure}[!ht]
\centering
\includegraphics[scale=0.8,angle=0]{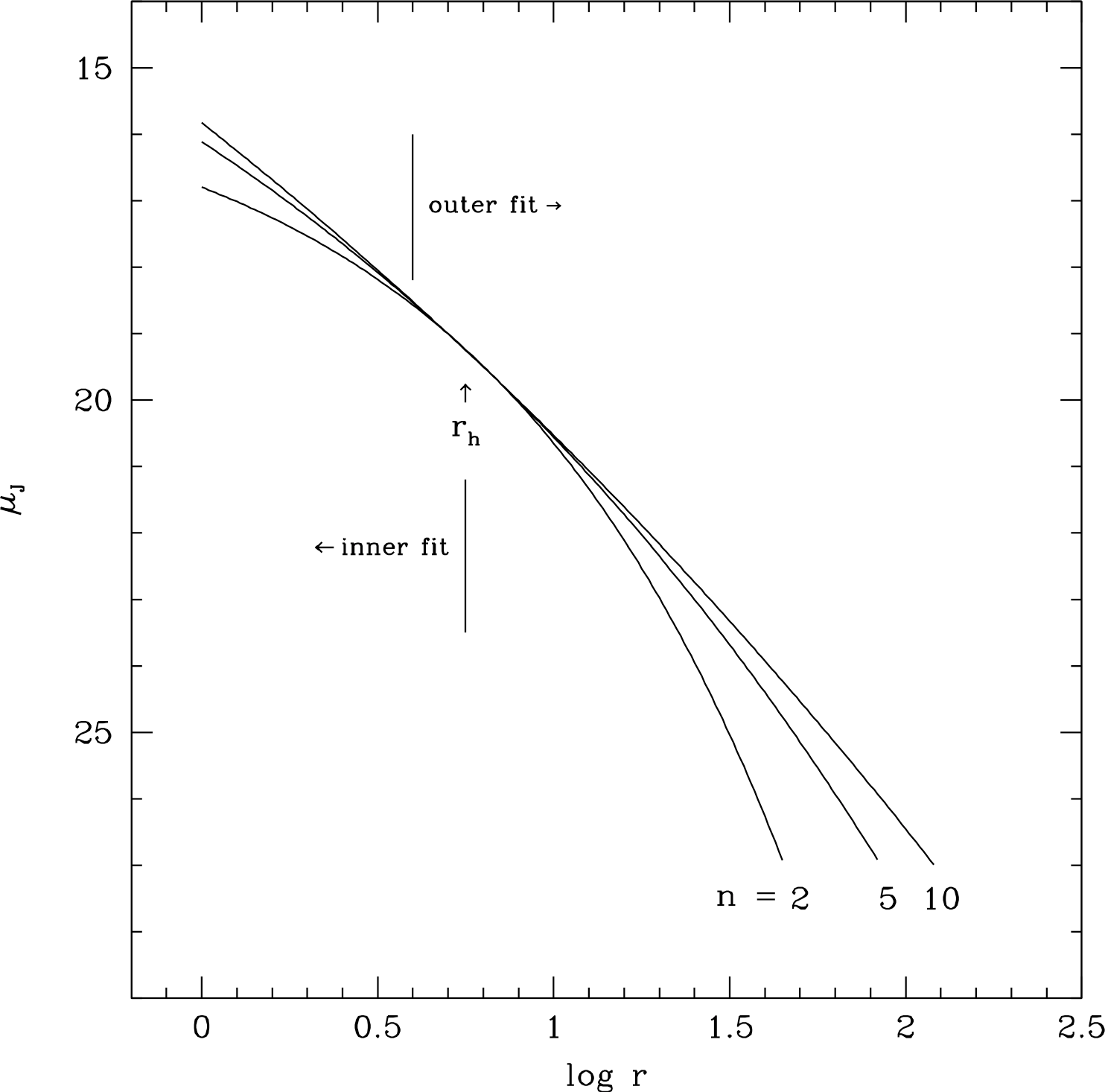}
\caption{\small The behavior of the S\'{e}rsic $r^{1/n}$ model $n$ index
for typical values of $\mu_e$ and $r_e$.  Lower $n$ provides more curvature
to a profile shape, particularly useful for fitting low luminosity
ellipticals and the core regions ($r$ less than the half-light radius,
$r_h$) of normal ellipticals.  However, the outer isophotes of most
ellipticals have shallower slopes (i.e., higher $n$ values) producing a
conflicting fitting process where lower scatter (e.g., greater weight) core
regions drive $n$ downward and shallower outer regions, but with higher
uncertainties, drive $n$ to higher values.  The regions for our inner and
outer fits are indicated with respect to the half-light radius, $r_h$.
}
\label{n_index}
\end{figure}

The success of the S\'{e}rsic $r^{1/n}$ model derives primarily from the
fact that it has an additional fitting parameter providing an extra degree
of freedom.  This immediate addresses the problem with the $r^{1/4}$ law in
the outer regions by supplying more flexibility to the fitting function at
large radii.  However, a difficulty for the S\'{e}rsic $r^{1/n}$ model is
that the $n$ parameter is sensitive to both the inner and outer shape of a
galaxy profiles in a dependent fashion (see Graham \& Driver 2005 for a
full review of the characteristics of the S\'{e}rsic $r^{1/n}$ model).  As
can be seen in Figure \ref{n_index}, the $n$ index drives the inner and
outer profile fit upward (brighter) in surface brightness for higher values
of $n$ (higher $n$ equals more concentration of central light).  Normal PSF
and core effects (e.g., coreless versus core ellipticals, Kormendy \etal
2009) would serve to drive $n$ downward, while extended halos would drive
$n$ upward.  Thus, there is no expectation that $n_{inner}$ values are the
same as $n_{outer}$ values.   An addition problem arises in that, when
fitting the entire profile, inner data points have smaller errors (plus
more numerous data points as ellipse fitting in high luminosity regions are
more compressed) and, therefore, are given greater weight to most fitting
algorithms.

For comparison, all 311 ellipticals were fit with a S\'{e}rsic $r^{1/n}$
model from the inner 5 arcsecs out to the half-life radius ($r_h$, this
typically corresponds to a surface brightness of $\mu_J = 20$).  This inner
fit sample is than compared to a sample which is only fit from the point
where the surface brightness profile becomes $r^{1/4}$ in shape outward
(this was between 3 and 5 kpcs) to the outer most data points.  All the
fits use the isophote errors (mostly the error in the sky value) to weight
the data points.  Note that $n$ values above 10 are effectively identical
as their differences are asymptotically smaller for higher $n$.

\begin{figure}[!ht]
\centering
\includegraphics[scale=0.8,angle=0]{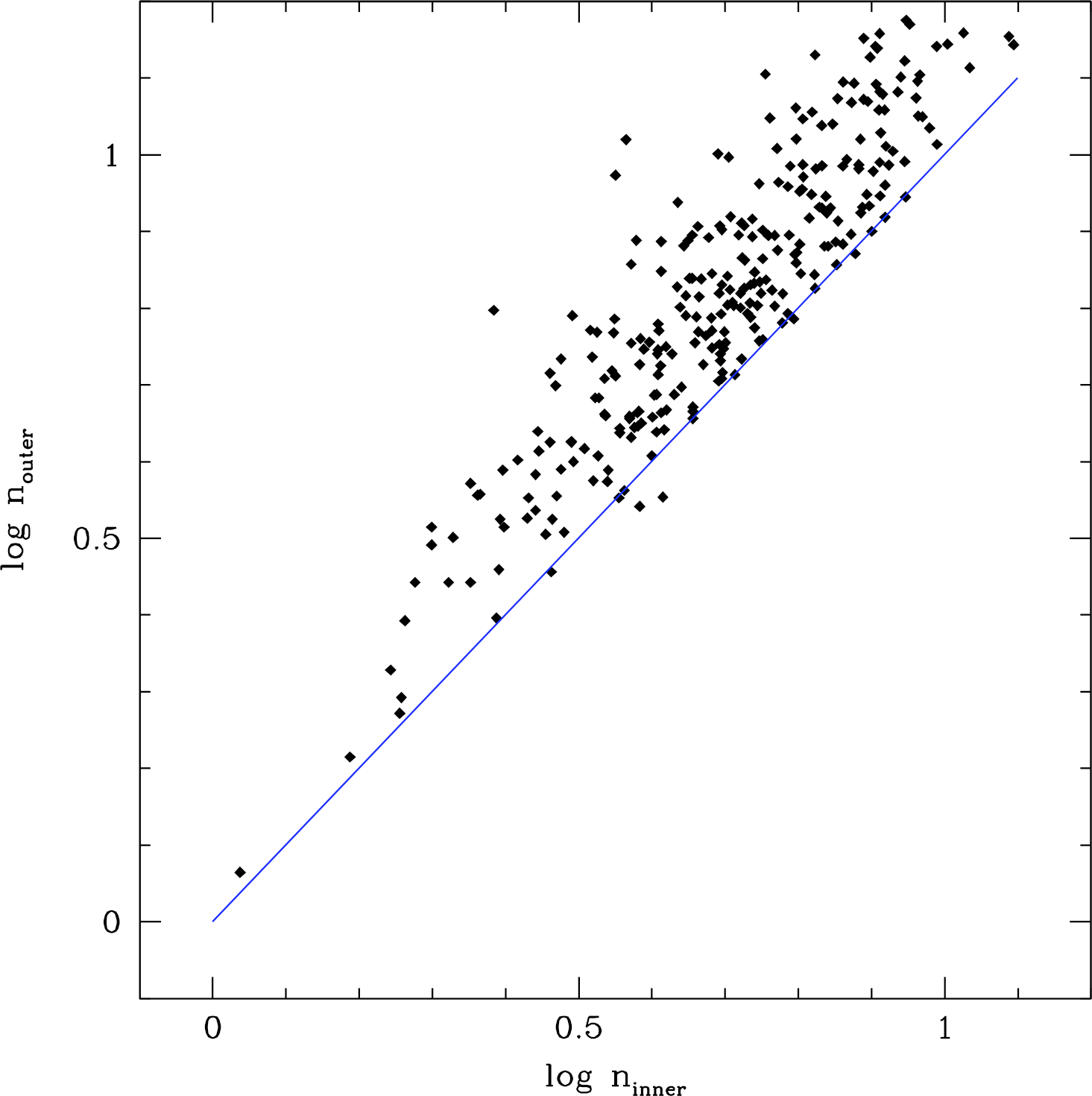}
\caption{\small The different S\'{e}rsic $r^{1/n}$ model $n$ values obtained by
fitting on the inner surface brightness profile (from 2 arcsecs to the
Holmberg radius, 22 $J$ mag arcsecs$^{-2}$) versus outer fits (from the
$r^{1/4}$ region to the outermost isophotes).  The shallower halos drive
the $n$ index to 80\% larger, on average, from the steeper core fits.  This
effect makes the S\'{e}rsic $r^{1/n}$ model ineffective as a universal description of
the full luminosity density profile of an elliptical.
}
\label{inner_outer}
\end{figure}

Unsurprisingly, the inner fit sample displays decreased $r_e$ (by 60\%) and
brighter $\mu_e$ (by 70\%, on average) compared to fits made to the 
halo (i.e., the inner and outer regions are not fit by the same model).  As
seen in Figure \ref{inner_outer}, the $n$ index is smaller for interior
fits by an average of 80\%.  This result is also independent of the fitting
constraints, for fits made to the entire galaxy profile simply resulted in
S\'{e}rsic $r^{1/n}$ model parameters identical to the inner fits as the
outer data points had greater photometric errors and were given less weight
by the fitting algorithms.

\begin{figure}[!ht]
\centering
\includegraphics[scale=0.8,angle=0]{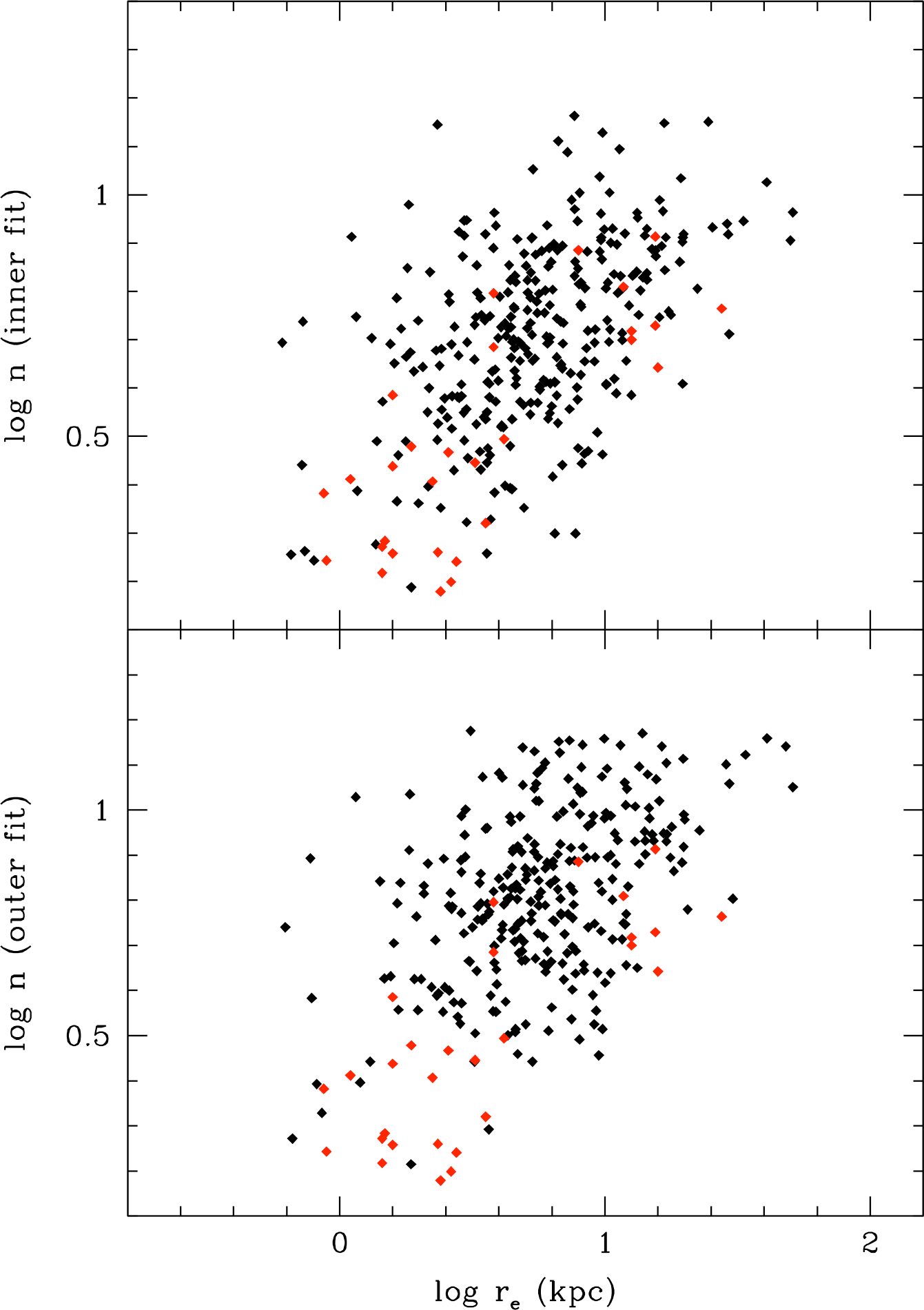}
\caption{\small The effective radius - S\'{e}rsic $r^{1/n}$ model $n$ index scaling
relation for $n$ values determined from inner fits (top panel) versus outer fits
(bottom panel).  The typically shallower profiles for ellipticals drives
$n$ to larger values for outer fits.  While the correlation is still
evident, the scatter is much larger than for inner fits.  The Caon \etal
data is shown as red symbols, based on high resolution inner fits.
}
\label{inner_outer_scale}
\end{figure}

The systematically different $n$ values between inner and outer fits
implies that it is impossible to find a photometrically correct match to an
entire elliptical surface brightness profile with a single component
S\'{e}rsic $r^{1/n}$ model.  It should be noted that $n_{inner}$ is weakly
correlated with $n_{outer}$ in Figure \ref{inner_outer}, but the variance
is too great for a single component fit. The effect on scaling relations
can been seen in Figure \ref{inner_outer_scale}, the S\'{e}rsic $r^{1/n}$
model effective radius ($r_e$) versus the $n$ index.  When the fits are
restricted to the inner regions, $n$ serves as a concentration index and
has a fair correlation with the effective radius, which is a measure of the
scale size of the galaxy (Trujillo \etal 2001).  However, when the fits are
restricted to the outer regions, the correlation with effective radius
degrades, $n$ serving as a measure of the shape of the halo, and becomes
very sensitive photometric errors from low surface brightness areas.

The method of fitting will also clearly influence the results.  For example, in
Figure \ref{inner_outer_scale}, the data from Caon \etal (1993) is shown and
clearly agrees with the inner fit distribution (although the correlation is
less evident than for the Caon \etal data).  But, the Caon \etal data has
lower $n$ values than those deduced for the outer fit sample, emphasizing
the importance of $n$ as a concentration indicator for the core region of
galaxies (Graham \& Guzman 2004).  PSF effects are a concern with 2MASS
images, but the same difference in $n_{inner}$ versus outer $n_{outer}$ is
evident even when the inner cutoff for the fit is varied.

\begin{figure}[!ht]
\centering
\includegraphics[scale=0.8,angle=0]{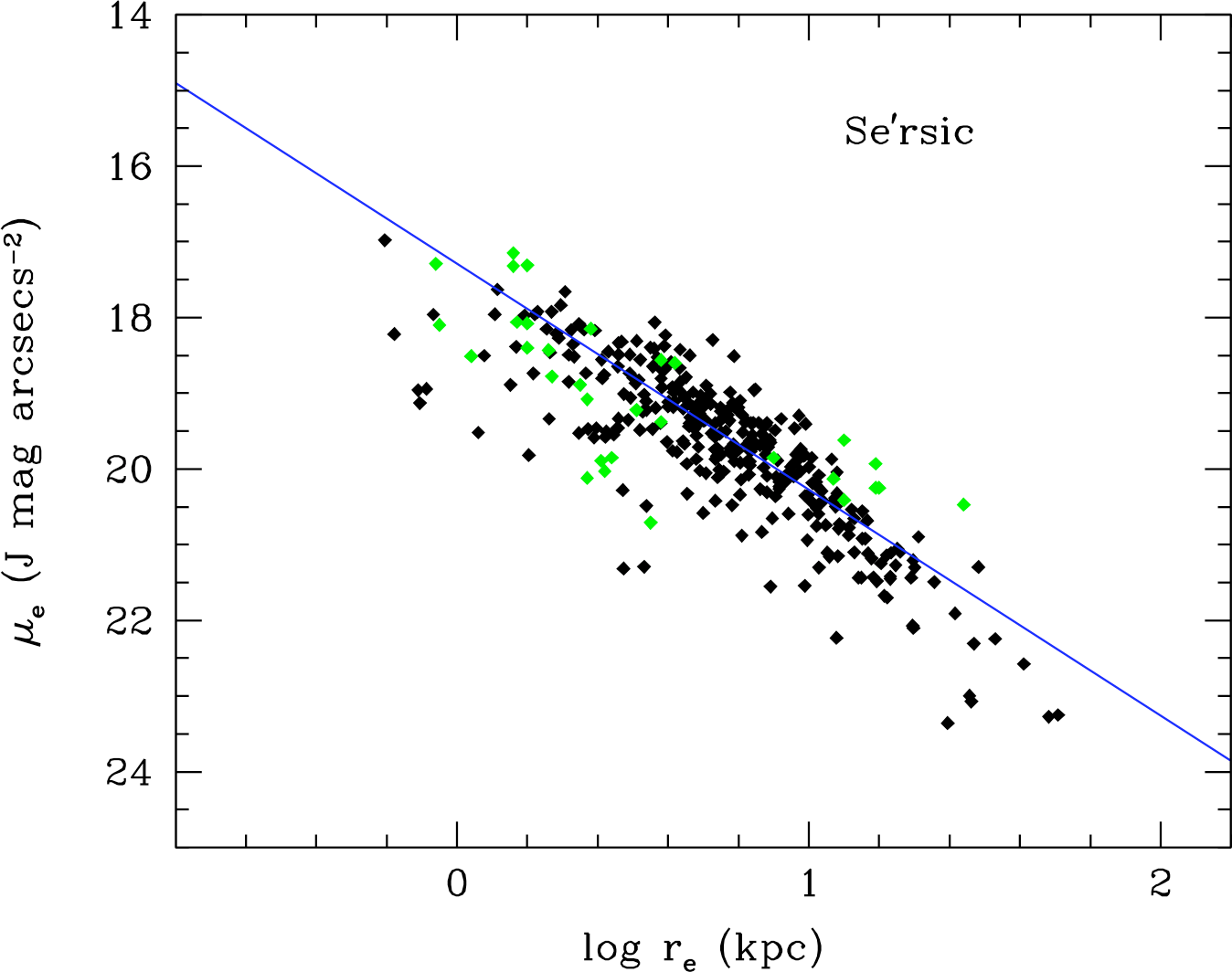}
\caption{\small The correlation between effective radius ($r_e$) and
effective surface brightness ($\mu_e$) for S\'{e}rsic $r^{1/n}$ model fits.  The blue line
is a jackknife linear fit, resulting in a similar relation to the $r^{1/4}$
fits in Figure \ref{re_se}.  The green symbols are the data from Caon \etal
corrected for color.
}
\label{auto_fit}
\end{figure}

For the rest of the analysis in this paper, the S\'{e}rsic $r^{1/n}$ model
is constrained to overweight the outer regions during fitting by
restricting the fit to only those points from the midpoint of the $r^{1/4}$
region to the halo.  In other words, the fitting is performed from the
radius where the inner isophotes becomes $r^{1/4}$ and continues outward,
weighted by surface brightness error for the outer points.  This inner
limit is always beyond 5 arcsecs, so PSF effects are negligible.  Other
inner radii were tested, for example, 1/4$r_h$, but all produced similar
results.  The resulting S\'{e}rsic $r^{1/n}$ parameters ($r_e$, $\mu_e$ and
$n$) are shown in Figures \ref{inner_outer_scale} and \ref{auto_fit}.

Immediately obvious from Figure \ref{auto_fit} is that $r_e$ and $\mu_e$
has a similar correlation as found from the $r^{1/4}$ fits.  The slopes are
identical, but the zeropoint is shifted by 0.3 mags fainter.  Even though
the $n$ index has a great deal of scatter, $\mu_e$ and $r_e$ are well
correlated and, again, the low scatter is assisted by the coupling of
$\mu_e$ and $r_e$.  While the additional free parameter increases the
quality of the fits for the S\'{e}rsic $r^{1/n}$ model (as measured by
$\chi^2$), in fact, there is no significant increase in the quality of the
$\mu_e$ versus $r_e$ diagram over $r^{1/4}$ fits.

The S\'{e}rsic $n$ parameter is weakly correlated with $r_e$ (Figure
\ref{inner_outer_scale} and $\mu_e$; however, the correlation is much
weaker than that found by Caon \etal (1993), shown as red symbols in
Figure \ref{inner_outer_scale}.  Much of this difference is, of course,
that Caon \etal focus on the use of $n$ as a central concentration
parameter, giving higher weight to the inner isophotes of a galaxy.  Our
procedure, to ignore inner isophotes, uses $n$ as a shape parameter for the
halo.  This appears to have the consequence of decoupling $n$ from
$r_e$ and $\mu_e$ since these latter parameters are more strongly
influenced by inner isophotes than outer ones (see below).

\begin{figure}[!ht]
\centering
\includegraphics[scale=1.2,angle=0]{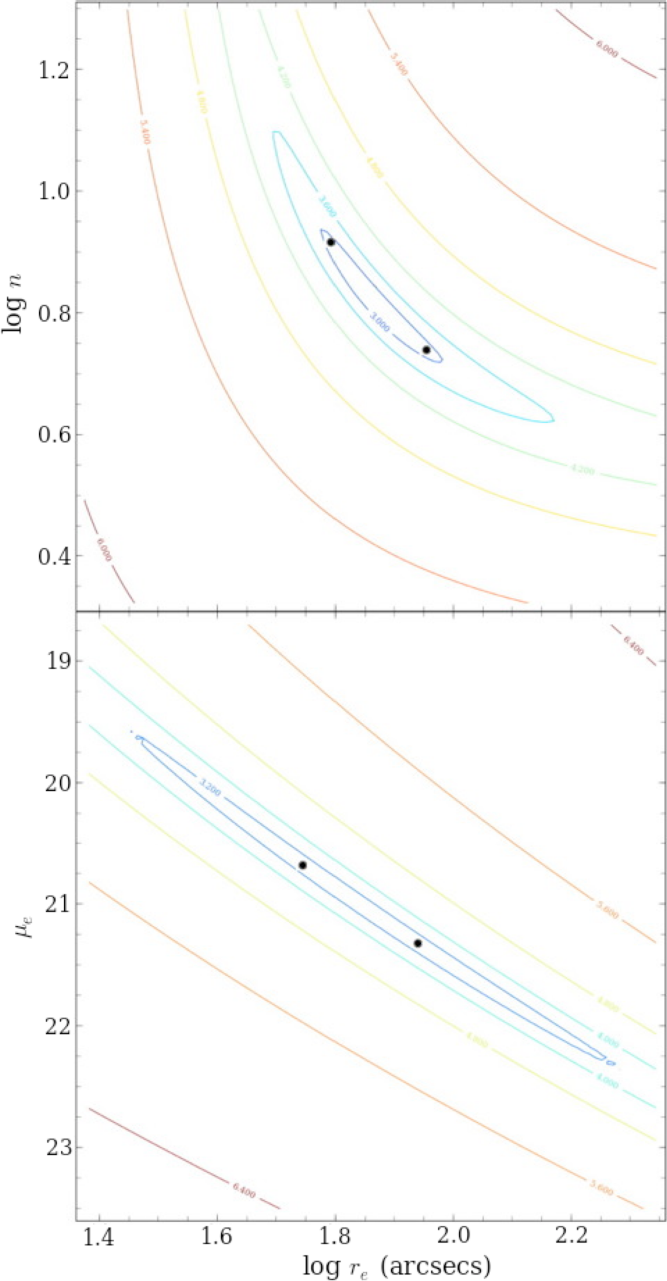}
\caption{\small The $\chi^2$ parameter space for S\'{e}rsic $r^{1/n}$ fits to NGC
7626 plotted against effective radius ($r_e$), effective surface brightness
($\mu_e$) and the $n$ index.  Contour lines corresponds to lines of
constant fit quality, $\chi^2$, the square of the difference between the
fit and the actual data.  The regions of best fit between $r_e$ and $\mu_e$
are long, narrow ellipses, meaning that there is a wide range of these
parameters that produce equally good fits.  Likewise, the $\chi^2$ contours
for the $n$ index display a non-linear coupling with $r_e$.  The two
indicated fits (black symbols) are the fits shown in Figure \ref{compare_sersic}.
}
\label{chi_grid}
\end{figure}

The weakness of the S\'{e}rsic $n$ parameter is also related to the large
variance in fit parameters for similar quality fits.  Figure \ref{chi_grid}
displays the $\chi^2$ space around a range of $r_e$, $\mu_e$ and $n$ values
for NGC 7626.  The $\chi^2$ test is not the optimal method for determining a
best fit to a surface brightness profile for it assumes that the errors in
the photometry are gaussian and random when, in fact, the errors at faint
light levels is dominated by systematics in the sky value (Schombert \&
Smith 2012).  However, it does have the advantage of simply comparing the
fit to the data as a measure of the total residual value, and a straight
forward weight by sky error can be applied to the outer isophotes.  There
is no attempt herein to assign a minimal $\chi^2$ value for an adequate
fit, merely to use $\chi^2$ for comparison between various fit parameters.

As can be seen in Figure \ref{chi_grid}, the $\chi^2$ determination for
each fit is very shallow, and the slope of the error ellipse is roughly
$\mu_e \propto -3.1 {\rm log} r_e$ compared to the correlation slope
(Figure \ref{auto_fit}, $\mu_e \propto -3.0 {\rm log} r_e$).  This means
that, like the $r^{1/4}$ law, small errors in $\mu_e$ and $r_e$ vary the
parameters along the correlation and errors in the fitting procedure work
to reinforce the relationship.  Likewise, small changes in $\mu_e$ and
$r_e$ also result in the $n$ index varying in a non-linear fashion (top
panel).

\begin{figure}[!ht]
\centering
\includegraphics[scale=0.8,angle=0]{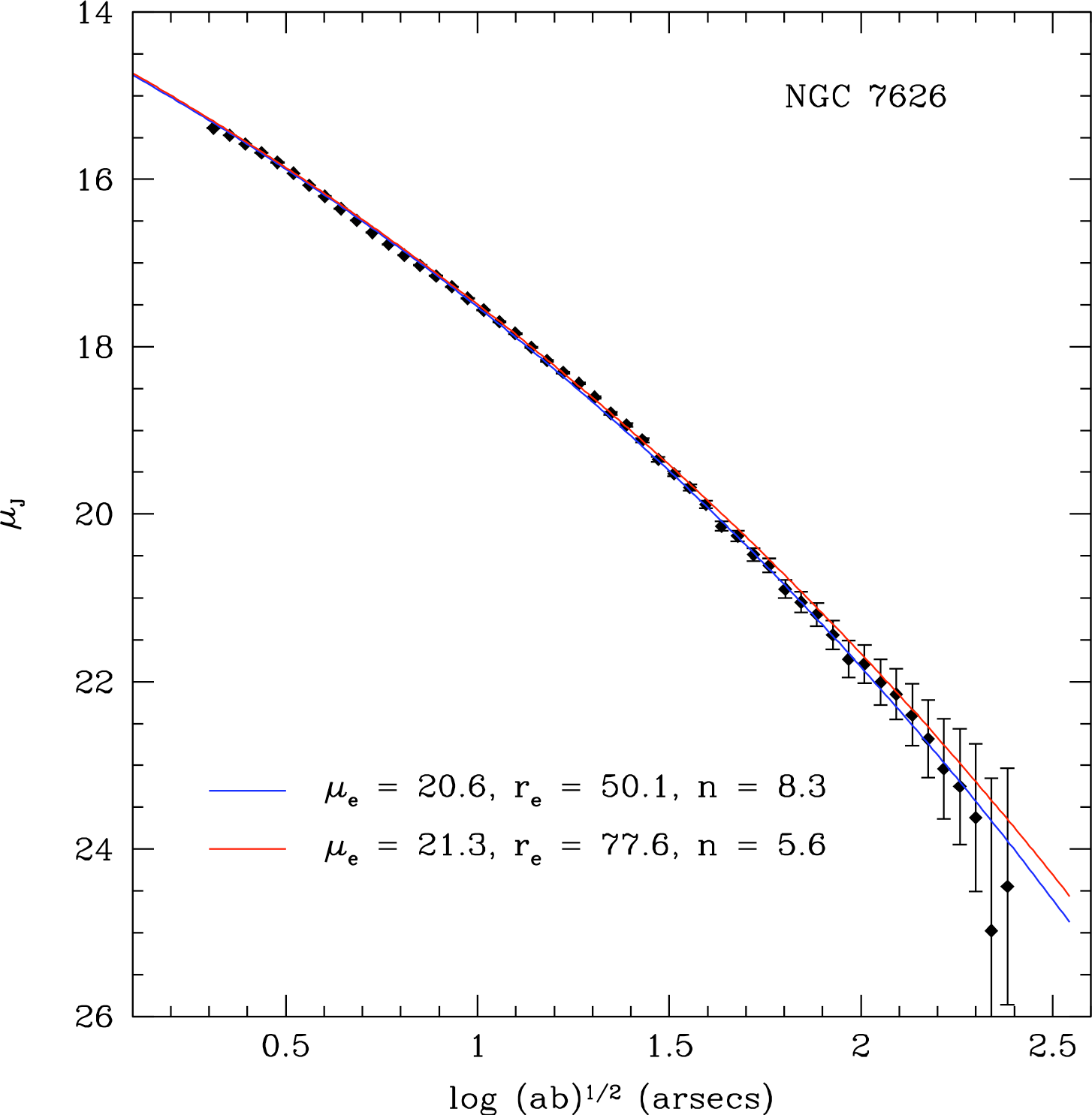}
\caption{\small The two S\'{e}rsic $r^{1/n}$ model fits shown in Figure
\ref{chi_grid} for NGC 7626.  While the blue fit has a slightly better
$\chi^2$ value, it is clear that, within the photometric errors, either fit
is equally valid.  Yet, the fit parameters ($\mu_e$, $r_e$ and $n$)
vary by 40\%.
}
\label{compare_sersic}
\end{figure}

In fact, a wide range of S\'{e}rsic parameters equally fit the profiles
within the errors of the data.  One example is found in Figure
\ref{compare_sersic}, where the two fits (indicated in Figure
\ref{chi_grid} as black symbols) are mapped onto the profile.  There is a
negligible difference in the quality of the fits, even though the fit
parameters vary by up to 40\%.  While the fit shown in blue is numerally
superior to the fit shown in red, within the errors of the photometry
either fit is equally valid.  Yet, there is a significant difference in the
fit parameters whose coupling allows for a much broader range in good fits
than one would find acceptable as an analytic technique.  The formal errors
on the fits do not take this coupling into account, and the true
uncertainties in the fitting parameters is much larger than quoted by many
authors.

\section{The Photometric Plane}

Following the technique outlined in Graham (2002), the best fit S\'{e}rsic
$r^{1/n}$ model parameters have been converted into `Photometric Plane`
(PP) values.  The PP is the photometric version of the Fundamental Plane,
first presented by Djorgovski \& Davis (1987).  For the PP, the $n$ index
serves as a proxy for velocity dispersion, which produces an immediate
observational advantage as photometric data is much easier to acquire than
spectroscopic values.  As our $n$ values are not as tightly tied to the
interior concentration of an elliptical, it was not immediately obvious
that the same photometric relations as found by Graham (2002) could be
extracted, although there is a weak connection between $n_{inner}$ and
$n_{outer}$.

\begin{figure}[!ht]
\centering
\includegraphics[scale=0.8,angle=0]{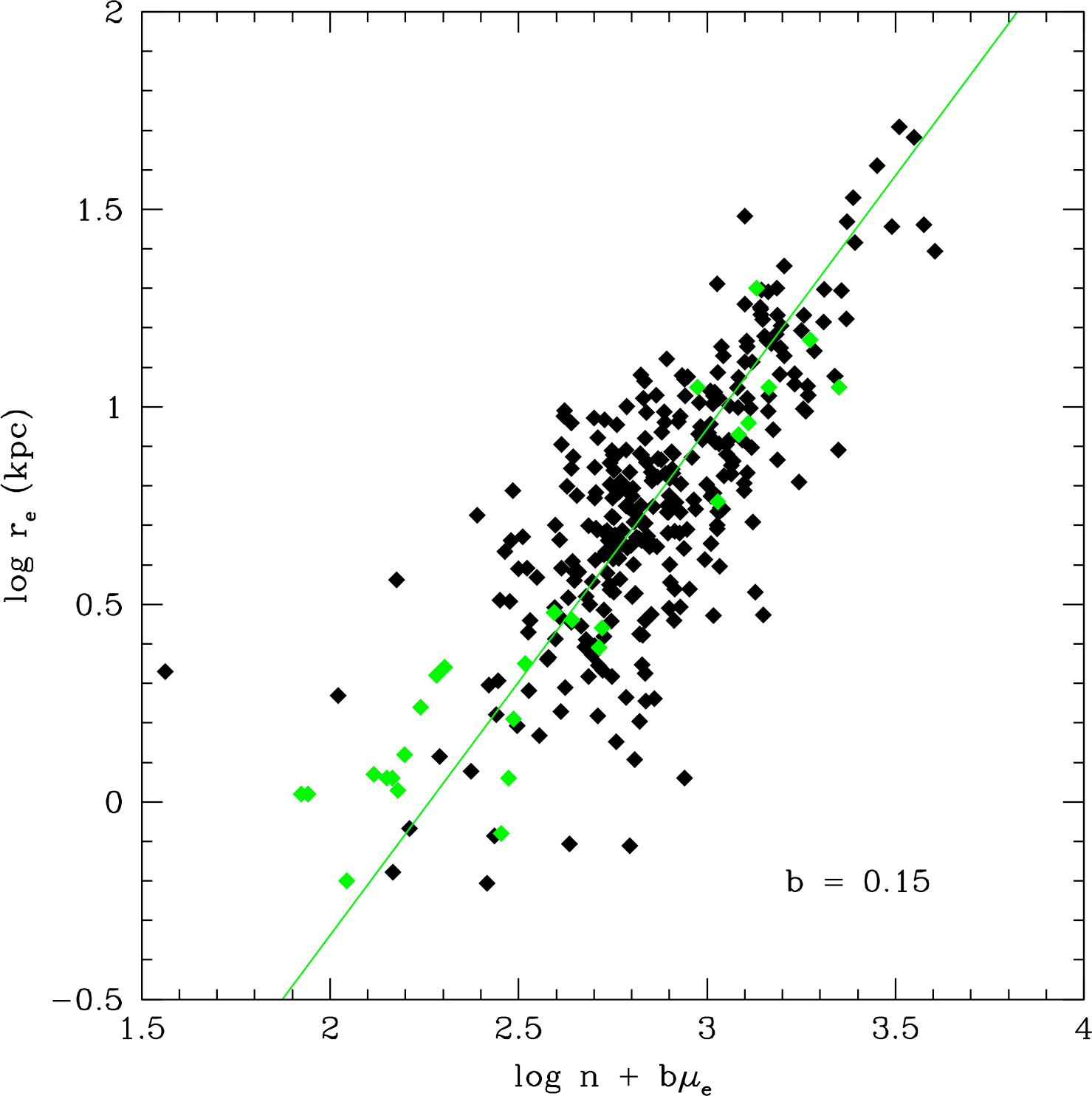}
\caption{\small The Photometric Plane, first proposed by Graham (2002), as
a correlation between scale length ($r_e$) and a linear combination of the
concentration index $n$ and luminosity density ($\mu_e$).  The green line
is a linear fit to the RMS minimized value of b=0.15.  Green data are the
original Caon \etal ellipticals used to formulate the original photometric
plane (corrected for a $B-J$ color).
}
\label{photo_plane}
\end{figure}

A best fit to PP values yields $r_e \propto n^{1.28 \pm 0.05}
\Sigma_e^{0.48 \pm 0.03}$ shown in Figure \ref{photo_plane}.  Also shown is
the Caon \etal data, corrected to $J$ with a $B-J=3.5$ color.  It is
somewhat surprisingly that the near-IR PP exists in our sample as our $n$
values are based on outer fits, whereas the original PP was based on $n$
values that were weighted towards inner regions.  Our difference in slope
for the $n$ index is primarily due to our different fitting methods, with
our $n$ values are larger, on average, than Caon \etal fits.

Interior fits are probably superior for the PP and discussions of
its meaning with respect to the specific entropy of an elliptical (the
Entropic Plane, Lima Neto \etal 1999) since the interior shape of an
elliptical more closely reflects the majority of the gravitational
potential.  In addition, the original motivation for the PP was the strong
correlation between galaxy velocity dispersion and $n$.  Our use of $n$ as
an outer profile shape parameter decouples that strong relationship, and
makes the PP less useful as the outer regions are strongly influenced by
post-formation processes.

Despite the differences in the $n$ values, the PP in Figure
\ref{photo_plane} displays a fair correlation.  Errors in $r_e$ and
$\mu_e$ track along the correlation, but most of the error budget is tied
to the uncertainties in $n$.  Typical 3$\sigma$ fit ranges are 0.14 in log $r_e$,
0.5 in $\mu_e$ and 0.1 in log $n$.  This results in an uncertainty in the
log $n$ + $b\mu_e$ axis of approximately 0.2, which would explain most of
the scatter in Figure \ref{photo_plane}.  As the near-IR bands quickly
redshift out of the observational windows, the near-IR PP is probably not
as useful as a distance indicator as the optically determined version.

\section{Scaling Relations}

The goal of structural analysis of ellipticals is to search for various
scaling relations (Graham 2011) that serve to outline a uniform sequence of
structural and luminosity (stellar mass) properties that ultimately
demonstrate structural homology and might be predicted by galaxy formation
models.  Before beginning this analysis, it should be noted that the sample
used in this study only outlines the upper end of the luminosity function
of ellipticals, those ellipticals brighter than $-$18 $B$ mag ($-$21.5 $J$
mag).  Only 7\% of our sample is faint enough to be classified as a low
mass or dwarf elliptical.  Thus, many of the issues outlined by Kormendy
\etal (2009) and Graham (2011) concerning the dichotomy of bright and faint
ellipticals are not addressed by our sample.

Perhaps the simplest structural parameters are the total luminosity (a
proxy for total stellar mass) and total galaxy size.  Whereas our technique
to use asymptotic functions, guided by a galaxies surface brightness
profile, produces highly reliable total magnitudes, this technique does not
lead to accurate total radii.  This is easy to see in the sense that small
errors in the outer profile will not significantly alter the luminosity
(as the light levels are lowest).  But, since the curve of growth flattens
at large radii, small errors in luminosity will lead to large variations
where one would define that last isophote.  Instead, the half-light radius
($r_h$) was selected because this has a lower uncertainty and it can be
compared to the effective radius as defined by the $r^{1/4}$ and S\'{e}rsic
$r^{1/n}$ functions.

\begin{figure}[!ht]
\centering
\includegraphics[scale=0.8,angle=0]{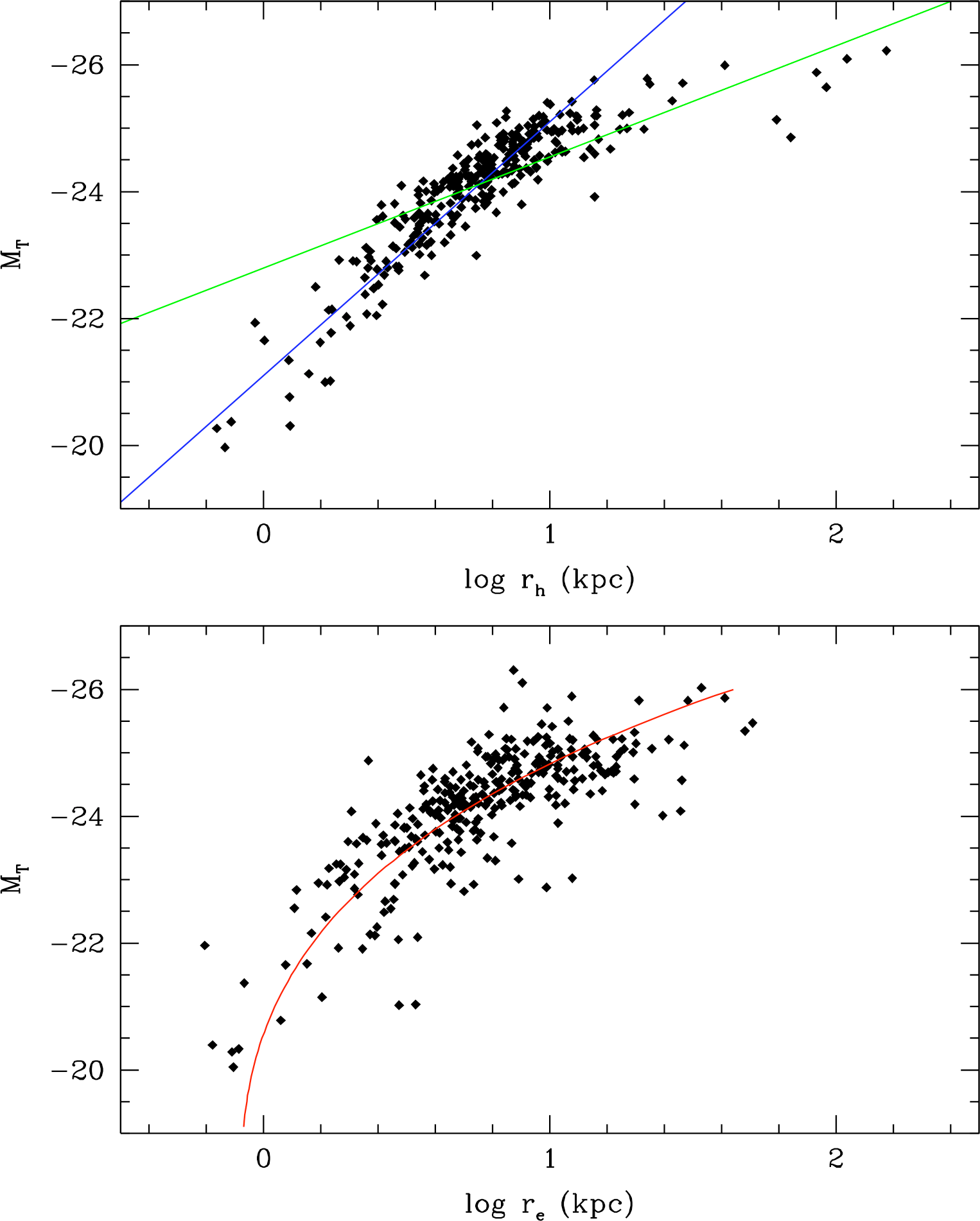}
\caption{\small The luminosity-radius relation using the empirically determined
half-light radius ($r_h$) and the effective radius ($r_e$) from S\'{e}rsic
$r^{1/n}$
model fits.  The blue line represents the $L \propto r^{0.7}$ (Kormendy
1977, Bernardi \etal 2007).  The green line represents $L \propto r^{1.6}$
(Strom \& Strom 1978).  The break at $M_T = -24 J$ was first discovered by
Schombert (1987).  The bottom panel displays luminosity versus effective
radius with the red line being the relationship from Graham \& Guzman (2003)
}
\label{half_rad}
\end{figure}

The luminosity-radius relation is shown in Figure \ref{half_rad}, where the
top panel displays the empirical half-light radius ($r_h$) which is the
point where the integrated light of the elliptical is 1/2 the total
luminosity ($M_T$).  The bottom panel displays the total luminosity versus
effective radius $r_e$ from S\'{e}rsic $r^{1/n}$ model fits.  This diagram is very
similar to the original luminosity-radius diagram published in Schombert
(1987, Figure 8)  based on $V$ photographic photometry.  As in the original
$V$ study, the correlation with radius appears to break into two separate
relationships for the bright and faint ends at approximate $M_T = -24$ $J$.
The break outlines the conflict between the relationship of $L \propto
r^{1.6}$, found by Strom \& Strom (1978) and the shallower relationship of
$L \propto r^{0.7}$, found by Kormendy (1977) and Bernardi \etal (2007).
Our original study measured the break at $-$20.5 in $V$, which corresponds
to $-$24 $J$ in Figure \ref{half_rad}.  The interpretation of this
effect is that bright ellipticals are more extended than their lower
luminosity counterparts, and is a prediction of dry merger scenarios
(Schombert 1987).

The difference between bright and faint ellipticals is less obvious in the
bottom panel, the luminosity versus effective radius ($r_e$) diagram.  Also
shown in that panel is the relationship outlined by Graham \& Guzman
(2003), corrected for a mean $B-J$ color of 3.5, a distinctly non-linear
relationship that connects bright and dwarf ellipticals.  While the data
agrees with the Graham \& Guzman relationship, the scatter is much larger
than the luminosity versus half-light radius diagram.  Data points farther
from the relationship are not poorer fits to the S\'{e}rsic $r^{1/n}$
model, so poor fitting does not explain the scatter, but probably reflects
the poor match between the S\'{e}rsic $r^{1/n}$ model and outer isophotes.

\begin{figure}[!ht]
\centering
\includegraphics[scale=0.8,angle=0]{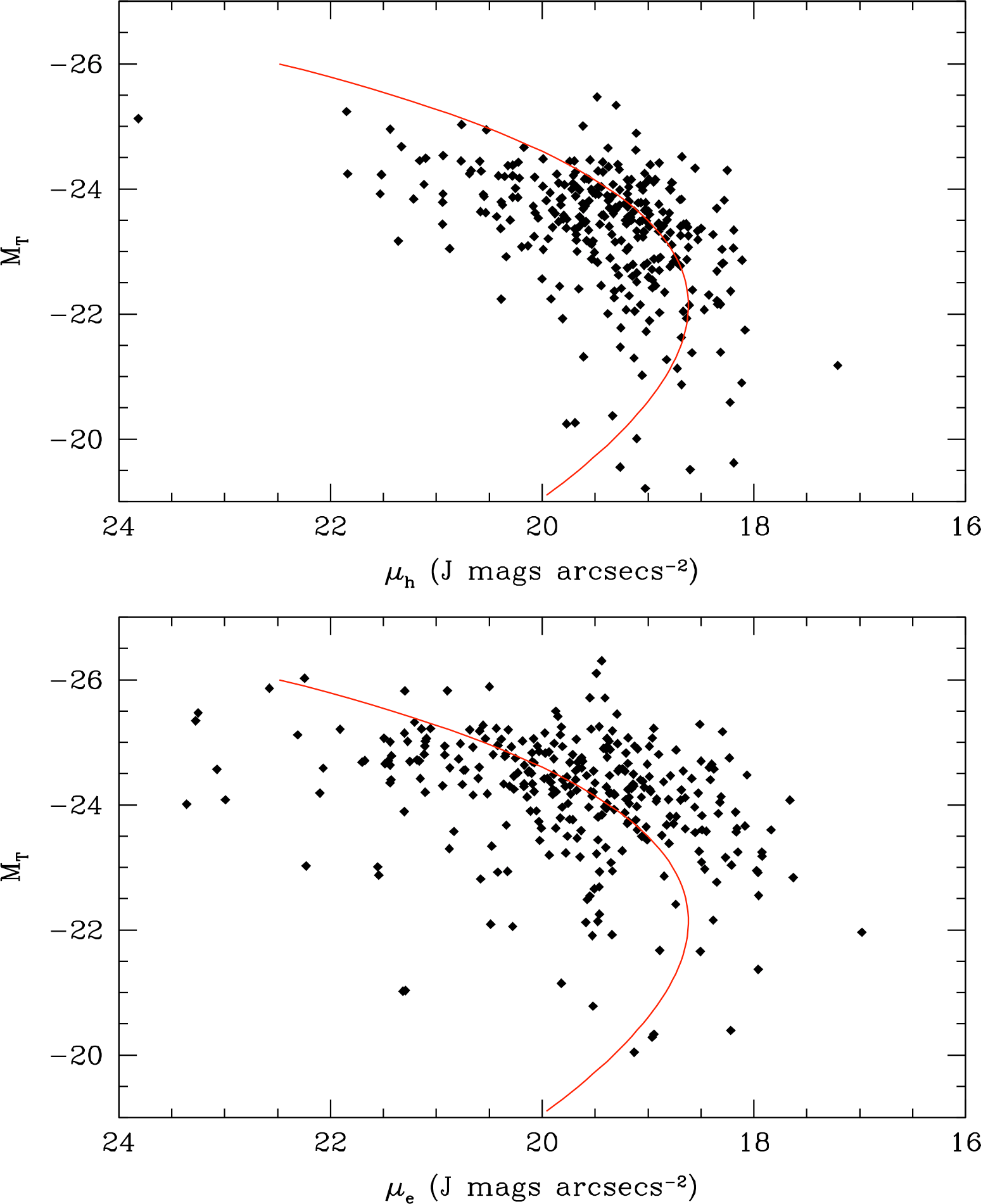}
\caption{\small Total luminosity versus half-light and effective surface
brightness.  The half-light surface brightness ($\mu_h$) is simply the
surface brightness of the galaxy profile at $r_h$.  The red line is the
relationship between luminosity and surface brightness found by Graham \&
Guzman (2003).
}
\label{eff_sfb}
\end{figure}

The luminosity versus half-light and effective surface brightness relation
is found in Figure \ref{eff_sfb}.  Here the half-light surface brightness
($\mu_h$) is defined as the surface brightness of the galaxy at the
half-light radius ($r_h$).  The effective surface brightness ($\mu_e$) is
derived from S\'{e}rsic fits.  There is no expectation that the
luminosity-surface brightness relation be linear (although a linear fit can
be made), certainly not by an extrapolation of the relationship of dwarf
ellipticals (Graham \& Guzman 2003), whose relationship is shown as the red
line in both plots.  While both distributions display similar shape, again,
the scatter in the empirically determined $\mu_h$ is less that $\mu_e$.
And, again as with the luminosity-radius diagram, there is a break at $M_T
= -24$ where the data maintains constant surface brightness for decreasing
luminosity.

\begin{figure}[!ht]
\centering
\includegraphics[scale=0.8,angle=0]{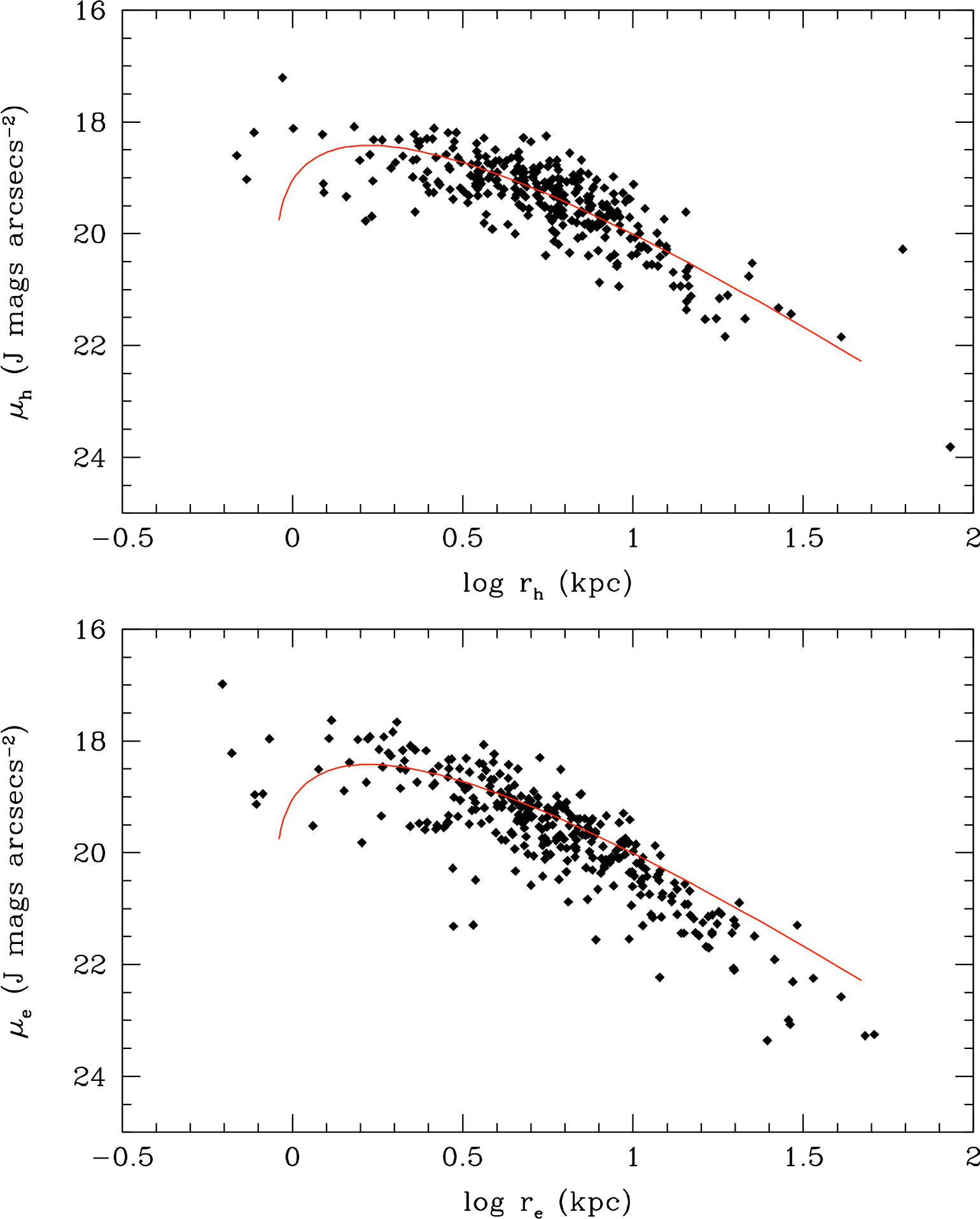}
\caption{\small Surface brightness versus scale length comparing empirical
half-light values with S\'{e}rsic fit values.  The red line is from Graham
\& Guzman (2003).  The apparent linear relation is, in fact, simply the
bright end of a more complicated relationship that decreases in effective
surface brightness for dwarf ellipticals (not shown).
}
\label{re_se_half}
\end{figure}

The correlation between surface brightness and scale length ($\mu_e$, $r_h$
and $r_e$) is shown in Figure \ref{re_se_half} (the same as Figure
\ref{auto_fit} for the S\'{e}rsic parameters).  As with the previous
diagrams, the scatter is less for the empirically determined half-light
radius ($r_h$), reflecting the added uncertainty induced by fitting
functions which are not necessarily an adequate description of the shape of
the galaxy's profile.  Previous work found this relationship to be linear
(Schombert 1987), but extensions to dwarf ellipticals (Graham \& Guzman
2003) finds that the correlation must drop in effective surface brightness
at small effective radii in order to make a continuous sequence from bright
to faint ellipticals.

As noted by Graham (2011), the non-linear relations for $\mu_e$ versus
luminosity, and $r_e$ versus luminosity, effectively guarantee that $\mu_e$
and log $r_e$ will be non-linear as well.  The Graham \& Guzman's color
corrected relationship for $\mu_e$ versus $r_e$ is shown in Figure
\ref{re_se_half} and, interestingly, fits the empirical $\mu_h$ versus
$r_h$ better than the S\'{e}rsic parameters.  The change from a linear
slope at large $r_e$ to a flattening relationship of constant $\mu_e$ at
small $r_e$ is well explained by the Graham \& Guzman curve.

\begin{figure}[!ht]
\centering
\includegraphics[scale=0.8,angle=0]{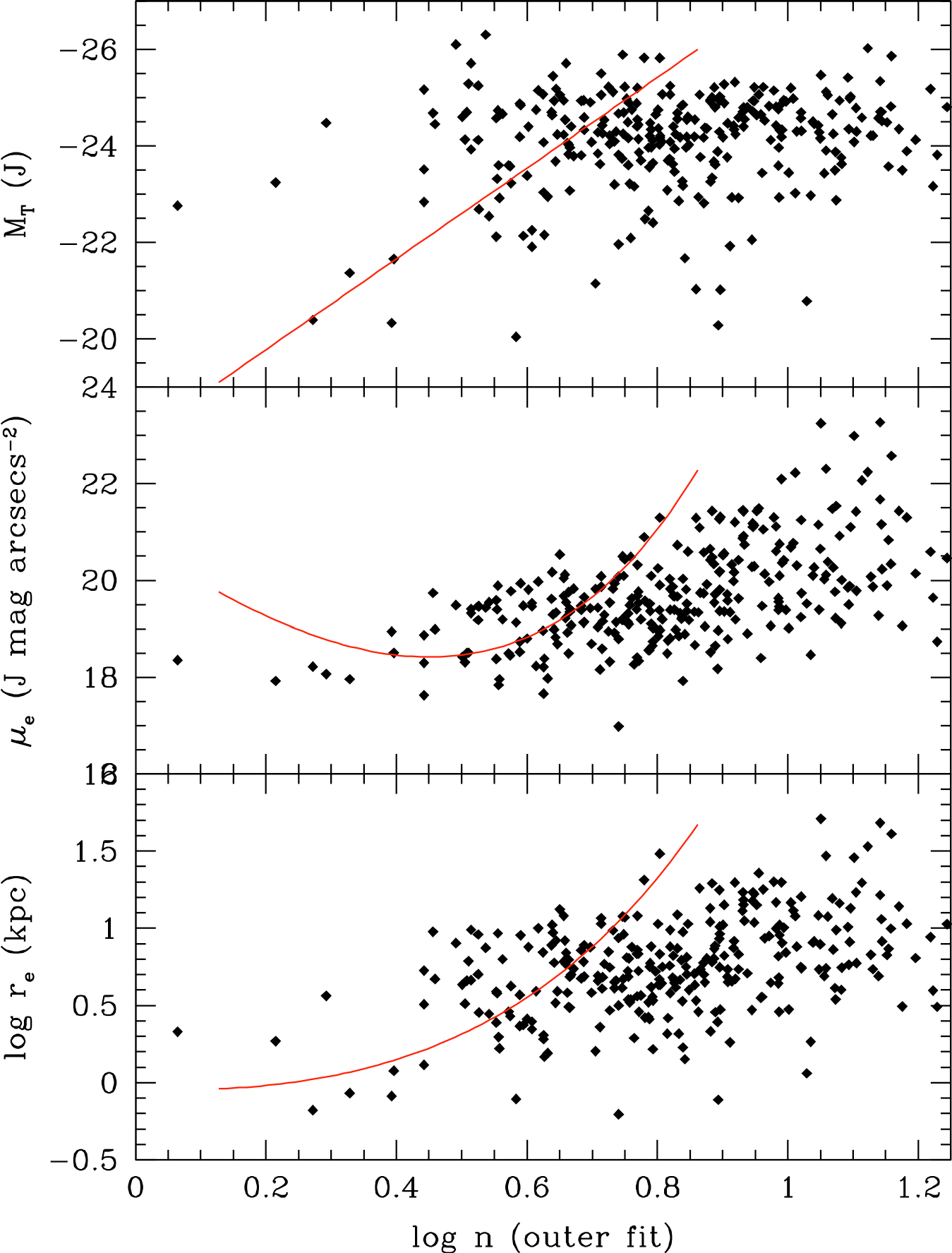}
\caption{\small The relationship between the S\'{e}rsic $n$ index and
$M_T$, $\mu_e$ and $r_e$.  The well defined relationships found by Graham
\& Guzman (2003, red curves) disappear when outer isophotes are used in the
fitting process.  There are mild trends of increasing $n$ with larger
galaxies (shallower profiles), but extracting useful structural information
in the halos of elliptical with the S\'{e}rsic $r^{1/n}$ model is lost.
}
\label{full_n}
\end{figure}

The remaining scaling relations between $M_T$, $\mu_e$, $r_e$ and the
concentration index $n$ are shown in Figure \ref{full_n}.  Unlike the well
defined correlations found by Graham \& Guzman (2003) (shown as red lines
in the Figure), the relationship between $n$ and the other photometric
parameters is practically non-existent.  There is a mild trend for
increasing $n$ with $r_e$ and $\mu_e$, but there is no relationship with
total luminosity. 

The lack of correlations is simply a strong statement on the nature of the
$n$ index in the context of the procedure for fitting a surface brightness
profile.  Early work (Trujillo \etal 2001) focused on using the $n$ index
as a measure of the central concentration of a galaxy.  This was achieved
by higher resolution imaging of galaxy cores than available from 2MASS
images, combined with a restriction of using data from the outer isophotes.
In addition, the fitting process weights the data by surface brightness,
automatically giving inner isophotes greater weight in the fits compared to
the outer isophotes (there are typically more isophotes in the bright
regions as well since the typical reduction scheme uses larger and larger
apertures in the fainter surface brightness regions).

\begin{figure}[!ht]
\centering
\includegraphics[scale=0.8,angle=0]{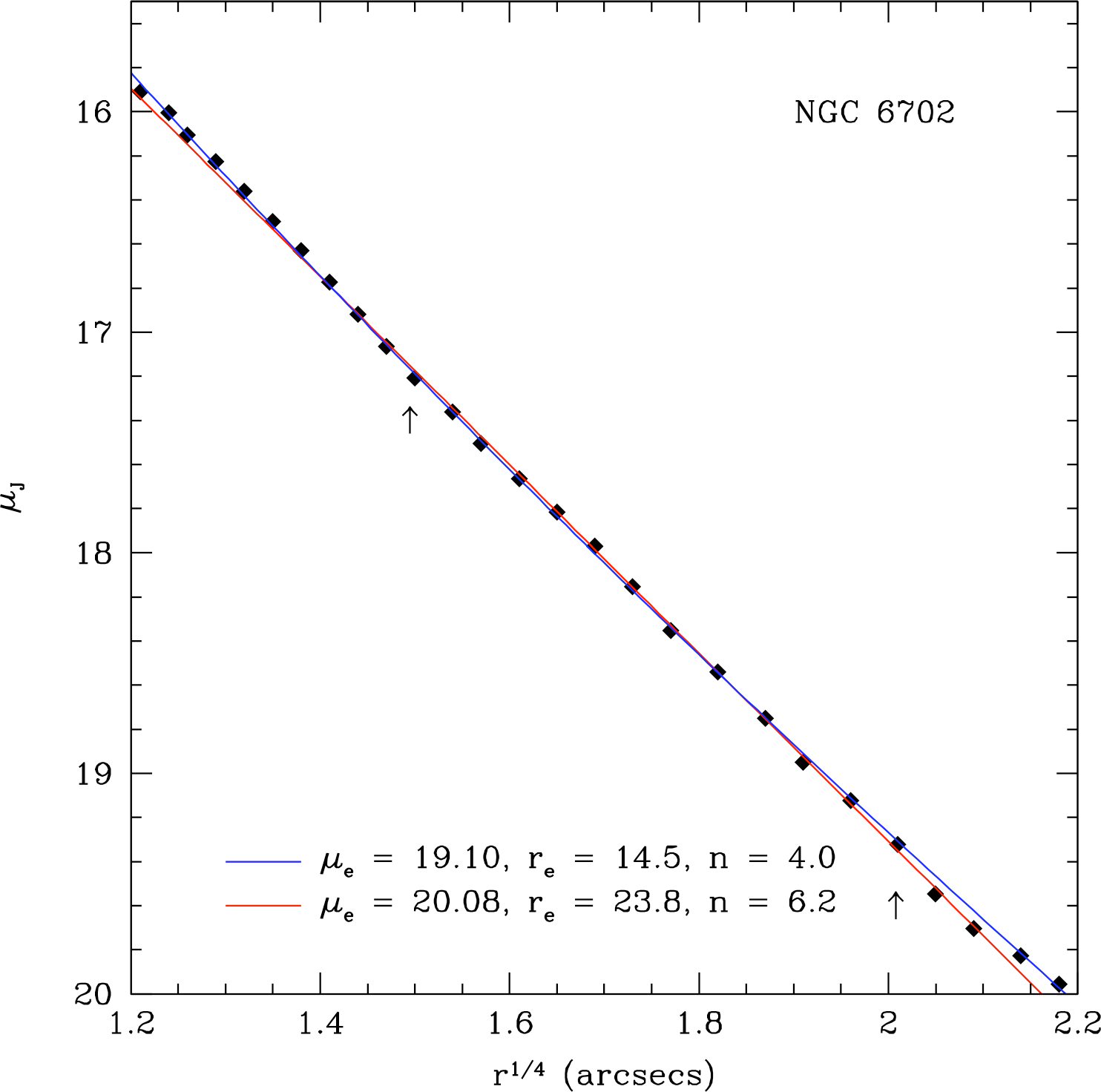}
\caption{\small Comparison of an $r^{1/4}$ fit versus S\'{e}rsic fit for
NGC 6702, a nearly perfect $r^{1/4}$ shaped profile (plotted in $r^{1/4}$
space for clarity such that the $r^{1/4}$ law is a straight line).  Even
when constrained to fit only the middle isophotes (indicated arrows), the
S\'{e}rsic $r^{1/n}$ model has too much coupling and flexibility to recover a correct
profile slope.
}
\label{sersic_vs_dev}
\end{figure}

The wide scatter in Figure \ref{full_n} underlies the intrinsic problem
with the S\'{e}rsic $r^{1/n}$ model for describing the halo of a
galaxy (that region beyond the half-light radius).  It simply does not have
the correct shape to capture the increasing shallow profile slope combined
with a sharp cutoff.  A clearer example can be found in Figure
\ref{sersic_vs_dev}.  Here the surface brightness profile of NGC 6702 is
plotted in $r^{1/4}$ space, and was selected for its large dynamic range in
surface brightness that appears $r^{1/4}$ (i.e., S\'{e}rsic $n=4$).  The
best $r^{1/4}$ (i.e., linear) fit is shown in blue, with a S\'{e}rsic $n$
index of 4 by definition.  However, the best S\'{e}rsic $r^{1/n}$ model fit
(between the two indicated limits) results in a formal fit $n$ index of
6.2.  The difference between the fits is negligible with very little
curvature at the faint and bright ends, yet a formal fit by a S\'{e}rsic
$r^{1/n}$ model wildly disagrees with a value of $n=4$, and decouples $r_e$
and $\mu_e$ from $n$.

In some sense, the S\'{e}rsic $r^{1/n}$ model is too flexible when
presented with data with a single power-law slope, but the very shallow
$\chi^2$ contours.  This results in a range of equally valid, but
ill-defined fits.  A range of value much larger than the formal errors
indicated by the fit algorithm.  When a flattened core structure is
present, than $n$ can become a well-defined measure of concentration, and
$r_e$ and $\mu_e$ parameterize the outer isophotes before the halo is
reached.  But, there is simply too much flexibility in the S\'{e}rsic
$r^{1/n}$ model in the outer regions of galaxies for it to be a unique
indicator of structure, even if the model fit itself accurately follows the
data.

While it is possible, using the S\'{e}rsic $r^{1/n}$ model, to find a set
of fit parameters that reproduces a majority of the inner or outer
isophotes, the uncertainty in the fit variables, as shown in Figure
\ref{chi_grid}, combined with the inability of the S\'{e}rsic $n$ index to
simultaneously follow the shape of the inner and outer portions of an
ellipticals profile, leads us to conclude that neither the $r^{1/4}$ law
nor the S\'{e}rsic $r^{1/n}$ model are adequate describers of the isophotes
of a typical elliptical brighter than $-$20 $J$ mag.  As once stated by a
famous galaxy photometrist, "It appears all fitting functions are simply
elaborate French curves to be inflicted on the data" (Oemler 1984).

\section{Summary}

The profiles of ellipticals have always held the greatest promise for
exposing underlying structural relations as they are uncluttered by
ongoing star formation, dust gas and irregular morphology.  Their
elliptical isophotes allows for the simplest reduction from 2D images to 1D
surface brightness profiles.  The analysis of these profiles has, in past,
used various mathematically relations (fitting functions) that, hopefully,
would have some analytic connection to underlying kinematics or, at least,
match predicted profiles from galaxy formation simulations.

In this work, using a large, uniform sample of ellipticals imaging in the
near-IR where the luminosity densities are the highest, the meaning and
usefulness of the two most common fitting functions, the $r^{1/4}$ law and
the S\'{e}rsic $r^{1/n}$ model have been examined.  The results are summarized
as the following:

\begin{description}

\item{(1)} The original discovery by Schombert (1986) is reinforced in that
the $r^{1/4}$ law only accurately describes the surface brightness profile
of an elliptical over a limited range of surface brightness and, in that
range, only for galaxies brighter than $-23 J$ mag.  The 1/4 power index is
an arbitrary for the power index as equally good fits are found for 1/5 or
1/3.

\item{(2)} With the above restrictions, the relationship between $\mu_e$
and $r_e$ is well defined across many wavelengths and studies; however, the
correlation is assisted by the strong coupling between the fit parameters
which serves to minimize the scatter and distort the true errors.

\item{(3)} The S\'{e}rsic $r^{1/n}$ model is a quantitative better fit to
elliptical profile, mostly due to its additional free parameter.  However,
there is no clear evidence that the shape of the outer isophotes is
correlated with the shape of the inner isophotes.  Therefore, the $n$ value
deduced from total profile fits will be heavily influenced by the lower
photometric error, and typically more numerous, inner isophote profile
points.

\item{(4)} Fits made to the inner portion (inside the half-light radius) of
a profile versus the outer portions (outside the $r^{1/4}$ region)
demonstrate that conflicting $n$ values are found.  The $n$ values for
outer fits are typically factors of two higher than inner fits, reflecting
the shallower profiles of the halo regions, and are only weakly correlated
with inner shape (see Figure \ref{inner_outer_scale}).

\item{(5)} Structural parameters extracted from the S\'{e}rsic $r^{1/n}$
model are reproducible between various studies; however, again, the meaning
of the fit parameters is highly distorted by the lack of uniqueness to the
fits due to strong coupling of the fit parameters.  The $chi^2$ space for
the fit parameters is wide and shallow, effectively allowing small
photometric errors to dominate the resulting fit values.  Nearly identical
fits are found with widely difference fit values (i.e., the fits are not
unique, see Figures \ref{compare_sersic} and \ref{sersic_vs_dev}).

\item{(6)} The S\'{e}rsic $r^{1/n}$ model photometric plane (Graham 2002) is
reproduced in the near-IR and using the $n$ fits to the outer isophotes.
It's slope and scatter are nearly identical to previous determinations,
even in light of the difficulty in applying the S\'{e}rsic $r^{1/n}$ model in a
coherent fashion.  Its linearity may be a reflection of the limited
luminosity range in our sample.

\item{(7)} Empirically determined values, such as half-light radius ($r_h$)
and surface brightness ($\mu_h$) are as, if not more, accurate compared to
S\'{e}rsic $r^{1/n}$ model fit parameters with respect to scaling relations
(luminosity versus $\mu$ or scale length).  All the scaling relations from
Graham \& Guzman (2003) are reproduced in the near-IR, with the exception of
correlations using the S\'{e}rsic $r^{1/n}$ model $n$ index.

\item{(8)} While none of the structural relations are linear, the bright
end of each sequence ($M_T < -23 J$ mag) are distinct from the fainter
galaxies, with a possible signature of mergers in the flattening of the
luminosity-radius correlation.  For the bright ellipticals, the scaling
relations were found to be $L \propto r^{0.8 \pm 0.1}$, $L \propto
\Sigma^{-0.5 \pm 0.1}$ and $\Sigma \propto r^{-1.5 \pm 0.1}$.  Although
these are only approximate power-laws inflicted on a much more complex
relationship between structure and luminosity.

\end{description}

If the ultimate goal is to relate some observed analytic function to
theoretical galaxy models, then the current suite of fitting functions are
inadequate.  As empirically defined parameters appear to have less scatter
(e.g., Figure \ref{half_rad}), then the best scheme to systematic describe
the shape of elliptical profiles is to allow the data to stand for
themselves.  In other words, to follow the prescription of Schombert (1987)
and build template profiles as a function of elliptical luminosity.  These
have the advantage of correctly containing all the curvature in 
structure that is not captured by a smooth analytic function, yet are more
stable than a spline fit in the sense that each template only has one
variable, the galaxy luminosity.  This technique will be the focus of our
next paper, and the application of this method to discover that ellipticals
are composed to two structural families (distinct from the core/cusp
problem).

\noindent Acknowledgements:
The software for this project was supported by NASA's AISR and ADP programs,
and images were obtain by the 2MASS project.  Conversations with Alister
Graham are gratefully acknowledged.

\pagebreak

\end{document}